\documentclass[aps,prl,twocolumn]{revtex4}
\usepackage{graphicx}
\usepackage{bm}
\usepackage{dsfont}

\newcommand{\ket}[1]{\mbox{$| #1 \rangle$}}

\begin{document}
\preprint{}
\title{
Entanglement and non-Markovianity of quantum evolutions}

\author{\'{A}ngel Rivas$^{1,2}$, Susana F. Huelga$^{1,2}$ and Martin B. Plenio$^{1,3}$}
\affiliation{$^1$Institut f\"{u}r Theoretische Physik, Universit\"{a}t Ulm, Ulm D-89069, Germany.\\
$^2$School of Physics, Astronomy \& Mathematics, University of Hertfordshire, Hatfield, AL10 9AB, United Kingdom.\\
$^3$ QOLS, The Blackett Laboratory, Imperial College London, London SW7 2BW, United Kingdom.
}

\date{\today}

\begin{abstract}
We address the problem of quantifying the non-Markovian character of quantum time-evolutions of general systems in contact with an environment. We introduce two different measures of non-Markovianity that exploit the specific traits of quantum correlations and are suitable for opposite experimental contexts. When complete tomographic knowledge about the evolution is available, our measure provides a necessary and sufficient condition to quantify strictly the non-Markovianity. In the opposite case, when no information whatsoever is available, we propose a sufficient condition for non-Markovianity. Remarkably, no optimization  procedure underlies our derivation, which greatly enhances the practical relevance of the proposed criteria.
\end{abstract}

\pacs{42.50.Lc,03.65.Yz,03.65.Ca,03.65.Ud}

\maketitle

The actual dynamics of any real open quantum system is
expected to deviate to some extent from the idealized Markovian
evolution that arises from the conditions of weak (or singular) coupling to a
memoryless reservoir \cite{RevKoss,BreuerPetruccione}. While quantum optics provides realizations
that are extremely well approximated by such an evolution, soft or condensed
matter systems evolve subject to conditions that are generally
unsuited to be treated within the Born-Markov framework
\cite{weiss}. This is particularly the case when considering
interacting many-body systems, where the subsystem's coupling strength
may be comparable to the coupling to the bath \cite{aki}. The exact
details of what makes a given quantum evolution non-Markovian may be
complicated, and in many cases, especially when thinking about many-body systems,
an accurate microscopic model of the system-bath
interaction may actually be unfeasible. It would therefore be very
useful to define some simple measure that captures, in some form,
the fact that the evolution departs from strict Markovianity. This problem was addressed by \cite{Wolf2} in the context of abstract quantum channels and, very recently, an optimization-based measure of non-Markovianity founded upon the behaviour of the trace distance under complete positive (CP) trace-preserving maps have been proposed in \cite{breuer}. 
%
\begin{figure}
\begin{center}
\includegraphics[width=0.45\textwidth]{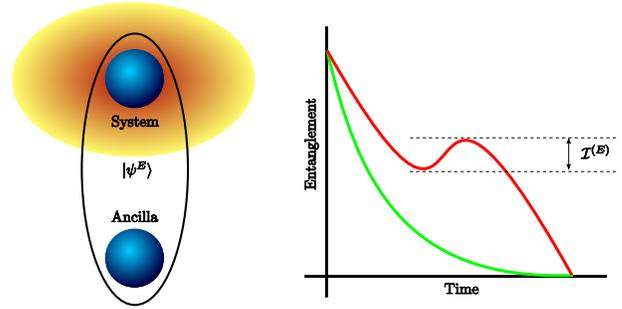}
\end{center}
\caption{Schematic illustration of the envisaged scenario to measure the non-Markovian character of an unspecified dynamical map.
An arbitrary quantum system, possibly multilevel and with an internal dynamics, is subject to the action of a local bath, depicted as a golden glow.
The system is initially prepared in a maximally entangled state $\ket{\Phi}$ with an ancilla which is kept shielded from the bath.
The green line represents some typical decay of the initial entanglement for a Markovian evolution while the red line
corresponds to a possible non-Markovian decay. In this case, the local action of the bath is no longer represented by a continuous family of CP propagators (\ref{complaw}) and the
entanglement between system and ancilla is no longer constrained to decrease monotonically. It is this deviation $\mathcal{I}^{(E)}$ that allows us to estimate the non-Markovianity of
the process despite ignoring any details about the evolution itself.}
\end{figure}

Formally, the dynamics of a quantum system given
by a family of trace-preserving maps $\mathcal{E}_\tau$ is called
Markovian if it defines a one-parameter semigroup of CP maps,
so that $\mathcal{E}_{\tau_1}\mathcal{E}_{\tau_2}=\mathcal{E}_{\tau_1+\tau_2}$.
With this definition, Markovian quantum processes are analogous to
their classical counterparts (see, for example,
\cite{bookSemiG,BreuerPetruccione}), the classical requirement of positivity being now replaced
by complete positivity as a result of the possible presence of genuine
quantum correlations (entanglement) with some extra system. The
structure of this kind of semigroups was analyzed in detail some
time ago \cite{Koss-Lind}, concluding that a quantum system will
undergo a Markovian dynamics provided that its evolution satisfies a
Master equation of the standard (Lindblad) form:
\begin{equation}\label{lindblad}
\frac{d\rho}{dt}=\mathcal{L}(\rho)=-i[H,\rho]+\sum_k\gamma_k\left(V_k\rho V_k^\dagger-\frac{1}{2}\{V_k^\dagger V_k,\rho\}\right).
\end{equation}
Here $H$ is a self-adjoint operator, $\gamma_k\geq0$, $\forall k$, and $\mathcal{L}$
is called the generator of the semigroup $\left(\mathcal{E}_\tau=e^{\mathcal{L}\tau}\right)$.
In analogy to the case of time dependent Hamiltonians in closed systems, the generators
can be also time dependent $\mathcal{L}_t$; then it is possible to prove that if and only if
$\mathcal{L}_t$ can be written in the standard form (\ref{lindblad}), with $H(t)$, $\gamma_k(t)\geq0$
and $V_k(t)$ potentially time dependent, the family of propagators $\mathcal{E}_{(t_2,t_1)}=\mathcal{T}\exp\left(\int^{t_2}_{t_1}\mathcal{L}_{\tau} d\tau\right)$ satisfying the composition law
\begin{equation}\label{complaw}
\mathcal{E}_{(t_2,t_0)}=\mathcal{E}_{(t_2,t_1)}\mathcal{E}_{(t_1,t_0)},
\end{equation}
for all $t_2\geq t_1\geq t_0\geq0$ are CP maps. This situation is sometimes referred to as time-inhomogeneous Markovian dynamics. Note that the condition (\ref{complaw}) is the quantum counterpart to the classic Chapman-Kolgomorov equation. \\
Conceptually, one could think of introducing a Markovianity measure
via some type of optimization problem, such as evaluating a quantity of the form
\[
\max_{\epsilon>0}\min_{\mathcal{E}^M}\Vert\mathcal{E}_{(t_0+\epsilon,t_0)}-\mathcal{E}_{(t_0+\epsilon,t_0)}^M\Vert,
\]
where $\Vert\cdot\Vert$ denotes some appropriate operator norm. The minimum is taken over the set of Markovian maps $\mathcal{E}^M$, and the maximum over final times deals with the time continuous dependence of the dynamical maps (note that $\mathcal{E}_{(t_0,t_0)}=\mathds{1}$ independently of $\mathcal{E}$, and $\mathds{1}$ is trivially Markovian). However, this quantity is hard to compute in practice due to the nonconvex
structure of the set of Markovian maps $\mathcal{E}^M$ \cite{Wolf2}.

In this Letter, we adopt a novel strategy and propose two possible ways to quantify the non-Markovian character of a quantum evolution which avoid the definition of an optimization problem. Our key element will be exploiting the specific behaviour of quantum correlations when a part of a composite system is subject to a local interaction that can be modeled as a trace-preserving CP map. This will allow us to give a necessary condition to measure deviations from Markovianity even when the actual form of the dynamics is completely unknown, although some non-Markovian evolutions may be undetected. A necessary and sufficient condition can be provided in the case when the dynamics is amenable to complete characterization, for instance, via quantum process tomography. Then, we are able to quantify strictly the
Markovian character of the evolution and deviations from strict Markovianity can be unambiguously characterized.

Let us consider first the case where we do not have any information about the dynamics of our system of interest, which 
we will consider initially to be a general, possibly composite 
$d$-dimensional quantum system. 
Our aim is to introduce a measure, that we denote by $\mathcal{I}^{(E)}$, that quantifies the deviation from Markovianity in the evolution of the system. For that, we will initially prepare
a maximally entangled state with an ancillary system which has to remain isolated from the decoherence sources, as illustrated in Fig. 1.
Since local trace-preserving CP maps do not increase the amount of entanglement
\cite{ShashMartin}, it is evident from the composition law (\ref{complaw}) that
the decay of the entanglement with an ancillary
system will be monotonically decreasing for Markovian evolutions. This fact also prevents the formation of loops in diagrams concurence vs purity as illustrated in \cite{ziman}.
However, if the evolution is non-Markovian, the requirement of strict monotonicity does no longer hold \cite{example},
as environmental correlations can lead to bipartite entanglement to be increased and decreased as a function of time (as exemplified by the red curve in Fig. 1).
Hence a conceptually simple way to quantify the degree of non-Markovianity of an unknown quantum evolution would be to
compute the amount of entanglement between system and ancilla at different times within a selected interval $[ t_0, t_\mathrm{max} ]$ and check for strict monotonic decrease of the quantum correlations. That is, for
$\Delta E=E[\rho_{SA}(t_0)]-E[\rho_{SA}(t_\mathrm{max})]$ (where $E$ denotes some entanglement measure) and some initial maximally entangled system-ancilla state, $|\Phi\rangle=\frac{1}{\sqrt{d}}\sum_{n=0}^{d-1}|n\rangle |n\rangle$,
$\rho_{SE}(0)=|\Phi\rangle\langle\Phi|$, we have
\[
\mathcal{I}^{(E)}=\int_{t_0}^{t_\mathrm{max}}\left|\frac{dE[\rho_{SA}(t)]}{dt}\right|dt-\Delta E,
\]
in such a way that if the evolution of the system is Markovian the derivative of
$E[\rho_{SA}(t)]$ is always negative and $\mathcal{I}^{(E)}=0$.\\
\begin{figure}
\begin{center}
\includegraphics[width=0.5\textwidth]{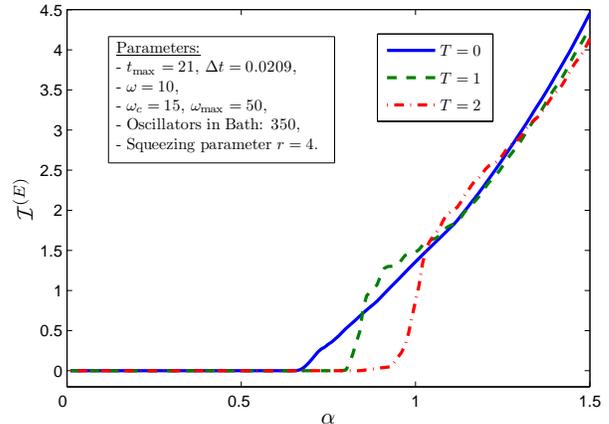}
\end{center}
\caption{Results of the simulation for the non-Markovianity as a function of the strength of the coupling $\alpha$ between a damped harmonic oscillator and a bath with Ohmic spectral density $J(\omega)=\alpha\omega e^{-\omega/\omega_c}$, for different temperatures. 
}
\end{figure}
Note that since the knowledge of the exact form of the dynamics is not necessary to measure $\mathcal{I}^{(E)}$, this method can be particularly useful in the study of infinite dimensional systems,
where the computation of the exact dynamical map is often difficult. For the sake of illustration, let us consider a single damped harmonic oscillator, with total system-bath Hamiltonian given by
\[
H=\omega a^\dagger a+ \sum_{j=1}^M\omega_ja^\dagger_j a_j+\sum_{j=1}^Mg_j(a^\dagger a_j + a a^\dagger_j),
\]
where $M$ is the number of oscillators in the bath, and we have assumed the validity of the rotating wave approximation (RWA). The bath is assumed to be initially in a thermal state $\rho_B=\exp\left(-H_B/T\right)/\mathrm{tr}\left[\exp\left(-H_B/T\right)\right]$, $H_B=\sum_{j=1}^M\omega_ja^\dagger_j a_j$,
and the system oscillator will be initially entangled with another oscillator, the ancilla, in a two-mode vacuum squeezed state $|\lambda\rangle=\sqrt{1-\lambda^2}\sum_{n=0}^\infty\lambda^n|n\rangle|n\rangle$,
where $\lambda=\tanh r$ and $r$ is the so-called squeezing parameter. Recall that this state is the most entangled Gaussian state at fixed mean energy $\bar{n}=\sinh^2 r$, and for
infinite squeezing $r\rightarrow\infty$ it approaches to the maximally entangled state. So it seems appropriate to apply our method, although the shape of the curves does not depend significatively on $r$. 
Since the Hamiltonian is quadratic in the annihilation and creation operators, it preserves the Gaussian character of the states and the amount of entanglement between system and ancilla can be computed easily by means of the logarithmic
negativity $E_N(\rho_{AB})=\log_2\Vert\rho_{AB}^{T_A}\Vert_1$, where $T_A$ denotes the partial transposition with respect to the subsystem $A$ and $\Vert \cdot\Vert_1$ the trace norm \cite{JensMartin}.
\begin{figure}
\begin{center}
\includegraphics[width=0.5\textwidth]{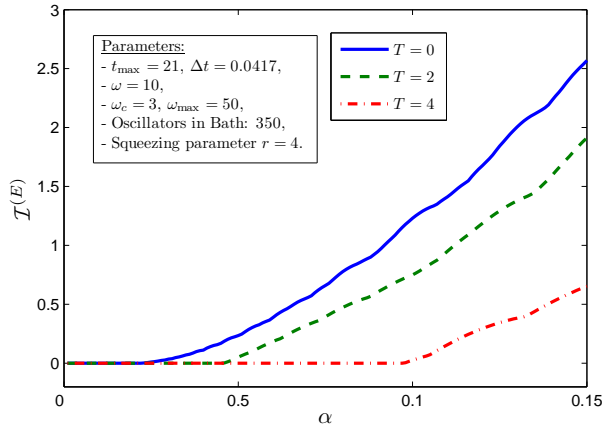}
\end{center}
\caption{The analogous to Fig. 2 for a super-Ohmic spectral density $J(\omega)=\alpha\omega^3 e^{-\omega/\omega_c}$, note the different values of the cut-off frequencies and the order of magnitude of $\alpha$.}
\end{figure}
To visualize the sensitivity of the proposed measure $\mathcal{I}^{(E)}$, two different spectral densities of the bath have been considered, as well as several initial temperatures.
In Fig. 2, the behaviour of $\mathcal{I}^{(E)}$ has been
plotted for an Ohmic spectral density with exponential cut-off $J(\omega)=\sum_{j=1}^Mg_j^2\delta(\omega-\omega_j)\rightarrow\alpha\omega e^{-\omega/\omega_c}$, where $\omega_c$ is the so-called cut-off frequency.
In order to make the analysis simpler, we focus only on the dynamical behaviour under different strengths of the coupling between system and bath. In the simulation, of course, the discrete and finite number of oscillators in the bath will affect the results, but in order to avoid this effect we have taken a number of oscillators
distributed in a range of frequencies (see details on the plots) such that our final time $t_{\mathrm{max}}$ is shorter than the recurrence time of the bath but large enough to capture
the general properties of the dynamics. It is clear from Fig. 2 that in the limit $\alpha\rightarrow0$ we approach a Markovian evolution independently of the value of $T$. This is the
well-known weak-coupling limit, whereas the on-set of deviations from Markovianity for stronger coupling strengths are $T$-dependent, so that the value of $\alpha$ for which
the measure $\mathcal{I}^{(E)}$ becomes nonzero increases with increasing temperature, the same happens with the value of $\mathcal{I}^{(E)}$ itself for $\alpha$ large enough.
An analogous situation is encountered in the presence of a super-Ohmic spectral density $J(\omega)=\alpha\omega^3 e^{-\omega/\omega_c}$, as shown in figure 3;  however, the specific value of $\mathcal{I}^{(E)}$ for given $T$ and $\alpha$ depends strongly on the bath spectral function.

There can be however non-Markovian quantum evolutions that remain undetected by the proposed measure, given that the requirement $\mathcal{I}^{(E)}>0$ is a sufficient condition for deviation from Markovianity. A necessary and sufficient condition can nevertheless be formulated if the specific form of the quantum evolution, as given by some dynamical
map $\mathcal{E}_{(t,t_0)}$, is amenable to exact reconstruction
between the some initial time $t_0$ and a final time $t$. This can in principle be done by means of
process tomography or perhaps resorting to a theoretical microscopic
model. Let us take our initial time as $t_0=0$ without loss of generality. Then, because of the
continuity of time, we can split the dynamical map as
\begin{equation}\label{semi-ncp}
\mathcal{E}_{(t+\epsilon,0)}=\mathcal{E}_{(t+\epsilon,t)}\mathcal{E}_{(t,0)}
\end{equation}
for any times $t$ and $\epsilon$. If the time evolution implemented by $\mathcal{E}_{(t,0)}$ is
Markovian, we have already mentioned that $\mathcal{E}_{(t_2,t_1)}$ is CP for any intermediate times $(t+\epsilon)\geq t_2\geq t_1\geq0$ (that is, $\mathcal{E}_{(t,0)}$ is infinitesimally divisible in the sense of \cite{Wolf2} for any $t$). However if and only if there exist times $t$ and $\epsilon$ such that $\mathcal{E}_{(t+\epsilon,t)}$ is not CP, the
dynamics will be non-Markovian. Note that this is the ultimate reason behind the possible increase of the system-ancilla entanglement at some
local times, which is the basis of our previous measure.
This partition can be extracted from the known dynamical map
$\mathcal{E}_{(t,0)}$ as a function of $t$ 
just by applying $\mathcal{E}^{-1}_{(t,0)}$ (which may not be a CP map) to the Eq. (\ref{semi-ncp}). Since
$\mathcal{E}_{(t\rightarrow0,0)}\rightarrow\mathds{1}$, for $t$
small enough $\mathcal{E}_{(t,0)}$ will be invertible, if for larger
times $\mathcal{E}_{(t,0)}$ is not invertible, we do not have enough
information to define $\mathcal{E}_{(t+\epsilon,t)}$ in an unequivocal way (this is one consequence of being blind to one part of the whole system). Then there are
several routes to follow depending on the nature of the singularities of $\mathcal{E}_{(t,0)}$; for example, strategies based on pseudoinverse maps have
been recently applied in a similar context \cite{pseudoInv}.

Being $|\Phi\rangle$ a maximally entangled state of our open system
and some ancillary one, because of the Choi-Jamio{\l}kowski
isomorphism \cite{Choi-Jam} $\mathcal{E}_{(t+\epsilon,t)}$ is CP if
and only if
$\left(\mathcal{E}_{(t+\epsilon,t)}\otimes\mathds{1}\right)|\Phi\rangle\langle\Phi|\geq0$.
Hence, given the trace-preserving property, we can take the following definition as a measure of the
non-CP character of $\mathcal{E}_{(t+\epsilon,t)}$,
\[
f_{NCP}(t+\epsilon,t)=\Vert\left(\mathcal{E}_{(t+\epsilon,t)}\otimes\mathds{1}\right)\left(|\Phi\rangle\langle\Phi|\right)\Vert_1.
\]
Therefore, $\mathcal{E}_{(t+\epsilon,t)}$ is CP if and only if
$f_{NCP}(t+\epsilon,t)=1$, otherwise $f_{NCP}(t+\epsilon,t)>1$.
Now
$f_{NCP}(t+\epsilon,t)$ will be the building block of our
measure of non-Markovianity. To construct it we leave $\epsilon$ to
be infinitesimal to define the (right) derivative of
$f_{NCP}(t+\epsilon,t)$:
\[
g(t)=\lim_{\epsilon\rightarrow0^+}\frac{f_{NCP}(t+\epsilon,t)-1}{\epsilon},
\]
noticing that $g(t)\geq0$, with $g(t)=0$ if and only if $\mathcal{E}_{(t+\epsilon,t)}$ is CP.
Therefore the integral
\[
\mathcal{I}=\int_0^\infty g(t)dt
\]
can be taken as a measure of non-Markovianity, and as long as $g(t)$ decreases fast enough (this will not be always the case) will be finite. Actually a normalized version of this measure can be $\mathcal{D}_{NM}=\frac{\mathcal{I}}{\mathcal{I}+1}$, in such a way that $\mathcal{D}_{NM}=0$ for $\mathcal{I}=0$ (i.e. Markovian
evolution) and $\mathcal{D}_{NM}\rightarrow1$ for $\mathcal{I}\rightarrow\infty$.


To illustrate the behaviour of $\mathcal{I}$, let us consider the dynamics of
one qubit modeled by a possibly non-Markovian differential master equation $\frac{d\rho}{dt}=\mathcal{L}_t(\rho)$ \cite{BreuerPetruccione}, since in the limit $\epsilon\rightarrow0$ the solution of this equation formally tends to $\mathcal{E}_{(t+\epsilon,t)}\rightarrow e^{\mathcal{L}_t\epsilon}$ \cite{reedsimon2}, we only need to expand it inside of trace norm up to first order to calculate $g(t)$,
\[
g(t)=\lim_{\epsilon\rightarrow0^+}\frac{\Vert\left[\mathds{1}+(\mathcal{L}_t\otimes\mathds{1})\epsilon\right]|\Phi\rangle\langle\Phi|\Vert_1-1}{\epsilon}.
\]
For instance, for a very simplified evolution of a qubit such as pure dephasing written like $\frac{d\rho}{dt}=\gamma(t)(\sigma_z\rho\sigma_z-\rho)$ we immediately obtain
\[
g(t)=\left\{\begin{array}{ll}
0 & \text{for } \gamma(t)\geq0 \\
-2\gamma(t) & \text{for } \gamma(t)<0
\end{array}\right.
\]
and finally
\begin{equation}\label{pd}
\mathcal{I}=-2\int_{\gamma(t)<0}\gamma(t)dt.
\end{equation}
Therefore, for this kind of evolution, $\mathcal{I}$ is proportional to the area of $\gamma(t)$
which is below zero. In particular, if the negative values of $\gamma(t)$ do not tend to zero fast enough $\mathcal{I}\rightarrow\infty$
[This is actually the case in the second example of \cite{breuer} where $\gamma(t)\sim\tan(t)$] \cite{compare}.

In summary, we have proposed two different approaches for the problem of quantifying non-Markovianity of general quantum evolutions; one is based on a sufficient condition whose evaluation does not require any prior knowledge of the quantum evolution itself, and the other provides a measure which quantify strictly the non-Markovianity, provided that the structural form of the dynamical map is known. Remarkably, the evaluation of the proposed measures does not require solving an optimization problem and would be suitable for providing information on the deviations from Markovianity in the experimental implementation of effective spin models using controllable systems, as for instance, trapped ions \cite{porras}. This type of experiments can prove extremely valuable in the subsequent formulation of detailed models of system-environment coupling in complex systems, both in condensed matter and, potentially, in some biological aggregates. These compounds have recently become experimentally probable at the femtosecond scale \cite{engel} but, given their complexity, no detailed microscopic models for the interaction with their surroundings are currently available.

\bigskip

We are grateful to A. Serafini for fruitful discussions and
A. Kossakowski for illuminating correspondence.
Financial support from the University of Hertfordshire,
the EU Integrated Project QAP, STREP action CORNER
and the Humboldt Foundation is gratefully acknowledged.



\begin{thebibliography}{00}
\bibitem{example}
A very simple example is a qubit interacting via the Hamiltonian $H=|01\rangle\langle10|+|10\rangle\langle01|$ with another qubit playing the role of the environment, i.e. the time evolution will swap back and forth with the evironmental qubit. Evidently this will periodically increase and decrease the entanglement between an ancillary qubit and the system.
\bibitem{compare} A similar result can be obtained for spontaneous emission and by using the Lorentzian spectral density for the Jaynes-Cummings model we would obtain a behaviour in total agreement with the results in Fig. 2 in \cite{breuer}.
\end{thebibliography}
\end{document}